\pdfoutput=1
\documentclass[10pt,aps,prb,amsmath,amssymb,twocolumn,letterpaper,
         citeautoscript,showpacs,raggedbottom,balancelastpage]{revtex4-1}
\usepackage[usenames,dvipsnames]{color}
\usepackage[pdftex,
  breaklinks,             %
  bookmarks = false, %
  pdfpagemode   = UseNone,%
  pdfstartview 	= FitH,		%
  pdfstartpage  = 1,      %
  colorlinks    = true,   %
]{hyperref}
\hypersetup{
    linkcolor=RubineRed,          
    citecolor=ForestGreen,        
    filecolor=Mulberry,      
    urlcolor=RoyalBlue           
}
\usepackage{graphicx}
\usepackage[protrusion=true,expansion=true,final]{microtype}
\usepackage{braket}
\usepackage{xspace}

\def\nmf{Na$_3$MnF$_6$\xspace}
\def\plow{[$P2_1/n$]$_1$\xspace}
\def\phigh{[$P2_1/n$]$_2$\xspace}



\begin{document}

\title{Microscopic origin of pressure-induced isosymmetric 
transitions in fluoromanganate cryolites}

\author{Nenian Charles}
   \email{neniancharles@drexel.edu}
	\affiliation{Department of Materials Science \& Engineering,\!
	Drexel University,\! Philadelphia,\! PA 19104,\! USA}%
\author{James M.\ Rondinelli}
  \email{jrondinelli@coe.drexel.edu}
	\affiliation{Department of Materials Science \& Engineering,\!
	Drexel University,\! Philadelphia,\! PA 19104,\! USA}%
\date{\today}

\begin{abstract}
Using first-principles density functional theory calculations, we investigate the 
hydrostatic pressure-induced reorientation of the Mn--F Jahn-Teller bond axis 
in the fluoride cryolite Na$_3$MnF$_6$.  
We find a first-order isosymmetric transition occurs between 
crystallographically equivalent monoclinic structures at approximately 
2.15 GPa, consistent with earlier experimental studies. 
Analogous calculations for isostructural $3d^0$ Na$_3$ScF$_6$ show no 
evidence of a transition up to 6.82 GPa. 
Mode crystallography analyses of the pressure-dependent structures in 
the vicinity of the transition reveals a clear evolution of the Jahn-Teller bond 
distortions in cooperation with an asymmetrical stretching of 
the equatorial fluorine atoms in the MnF$_6$ octahedral units.
We identify a change in orbital occupancy of the $e_g$ manifold in the 
$3d^4$ Jahn-Teller active Mn(III) to be responsible for the transition, 
which stabilizes one monoclinic $P2_1/n$ variant over the other. 
\end{abstract}

\pacs{61.50.Ks, 74.62.Fj, 71.15.Mb}

\maketitle

\section{Introduction}
\begin{figure}[b]
\centering\vspace*{-12pt}
\includegraphics[width=0.86\columnwidth]{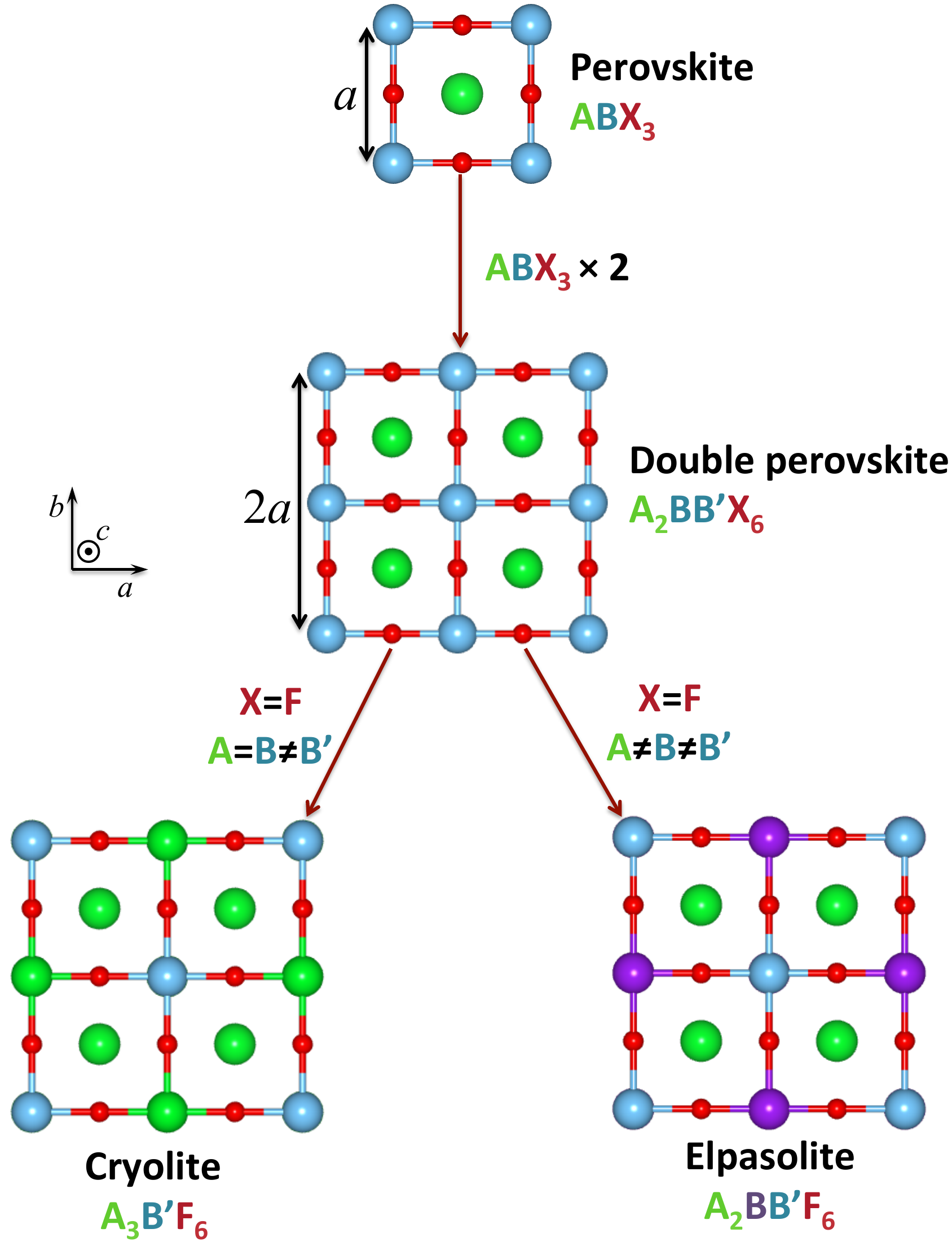}\vspace{-0.86\baselineskip}
\caption{%
Ideal structures of the double perovskite fluorides cryolite 
(A$_3$B$^\prime$F$_6$, left) and 
elpasolite (A$_2$BB$^\prime$F$_6$, right) as they relate to the 
cubic  ABX$_3$ perovskite structure (top).
} 
\label{fig:pervo_compare}
\end{figure}

Perovskite transition metal compounds with ABX$_3$ stoichiometry, 
where X is either oxygen or fluorine, have been extensively studied 
owing to their ability to support a rich set of physical phenomena  
ranging from multiferroisim to superconductivity \cite{Pinlac/Poeppelmeier:2011,Mulder/Benedek/Rondinelli/Fennie:2013,qiao2013impacts,martin2012multiferroic}. 
Many of these properties are a direct consequence of electron-lattice interactions 
across 
structural phase transitions owing to symmetry reductions, which alter 
chemical bonding pathways and electronic properties 
through distortions and rotations of BX$_6$ octahedra. 
Indeed, the physical properties of perovskites can be tailored by 
changing the bonding environments by manipulating the BX$_6$ size, shape, 
and connectivity\cite{RondinelliMayFreeland2012} using chemical 
pressure or epitaxial 
strain \cite{Imada/Fujimori/Tokura:1998,doi:10.1021/cm402063u,Schlom/Zurbuchen:2008,Martin/Ramesh_etal:2008}. 
Thus, for electronic function engineering purposes it is useful to understand 
the tendency of a perovskite to distort, or undergo a structural transitions, which can be 
estimated using the Goldschmidt tolerance factor \cite{Goldschmidt:1926}.
The tremendous success of perovskites has sparked 
interest in a closely related set of materials, 
the A$_2$BB'X$_6$ double perovskites with multiple cations that 
order among various planes in a periodic fashion (\autoref{fig:pervo_compare}).  
The additional compositional ordering degree of freedom not found in 
simple perovskites \cite{anderson1993b,Graham/Woodward:2010}
can be used to obtain desirable ferroic responses \cite{0953-8984-26-19-193201,young2013atomic},  
and explore correlated electron properties \cite{PhysRevLett.107.257201,PhysRevLett.85.2549,PhysRevLett.110.087203}.
It is important to note that while there are a multitude of studies on 
perovskites and double perovskites, \emph{i.e.}, most of the materials 
physics literature is keenly focused on X=O transition metal oxides \cite{Pinlac/Poeppelmeier:2011}, 
fluoride compounds, however, also show an affinity for functional 
electronic behavior, including mutiferroicity \cite{Ravez:1997,PhysRevB.74.020401,Scott/Blinc:2011,PhysRevB.89.104107}. 

For the case of rock salt ordered double perovskites with two possible 
B cations (B and B$^\prime$) with X\,=\,F (fluorine), 
these compounds are generally referred to  by one of the two prototypes 
illustrated in \autoref{fig:pervo_compare}. 
The first class being \emph{cryolite},  after the compound 
Na$_3$AlF$_6$ \cite{hawthorne1975refinement}.
This family contains compounds with [12]-coordinate A sites in the ideal structure, 
whereby the same cation on the A site is chemically identical to 
one at the octahedrally coordinated B site (A\,=\,B), while the 
B$^\prime$ cation remains unique.
This results in the stoichiometry A$_3$B$^\prime$F$_6$ for cryolite. 
The other prototype is \emph{elpasolite}, which is named for the 
mineral K$_2$NaAlF$_6$.\cite{moras1974crystal} 
It also refers to structures with 12-fold coordinated A cations, but 
unlike cryolite, the two octahedrally coordinated B sites are chemically 
distinct species (B and B$^\prime$).\cite{Flerov_etal:1998,Massa/Babel:1988}
Thus the stoichiometry for elpasolite is A$_2$BB$^\prime$F$_6$.
Similar to the transition metal perovksite oxides, the ground state properties of 
cryolites and elpasolites (double perovskite fluorides) are also strongly 
structure and symmetry dependent. 
The tendency for any double perovskite to distort is estimated using 
a generalized 
tolerance factor 
\[
t = \frac{\sqrt{2}(r_\mathrm{A} + r_\mathrm{X})}{(r_\mathrm{B} + r_{\mathrm{B}^\prime} + 2r_\mathrm{X})}\,,
\]
where $r_\mathrm{A}$, 
$r_\mathrm{B}$, 
$r_{\mathrm{B}^\prime}$, and 
$r_\mathrm{X}$ are the ionic radii of the different chemical species 
taking into account the appropriate coordination number and 
spin states approximated by the Shannon and Prewitt 
radii \cite{Shannon/Prewitt:1969, Shannon:1976}. 
In the case of double perovskite fluoride compounds  
($r_\mathrm{X}=r_\mathrm{F}$), 
the use of the Shannon radius for fluorine with a coordination number of 2 
($r_\mathrm{F}$ = 1.28.5\,\AA)  accurately reproduces observed 
B$^\prime$--F bond lengths \cite{Massa/Babel:1988}. 
For tolerance factor values in the range $0.91 \le t \le 1.00$, the cryolite and 
elpasolites tend to remain cubic---a much larger cubic stability field than 
found in ABX$_3$ perovskites,\cite{cook1991systematic} 
which are anticipated only to be cubic for $t = 1$. 
For the case where $t \textless 0.91$, the double perovskite fluorides 
also have a stronger affinity for octahedral rotations, and 
for $t \textgreater 1.00$, like the oxides, cryolite and elpasolite 
adopt hexagonal symmetries \cite{Massa/Babel:1988}. 

\begin{figure}
\centering
\includegraphics[width=0.99\columnwidth]{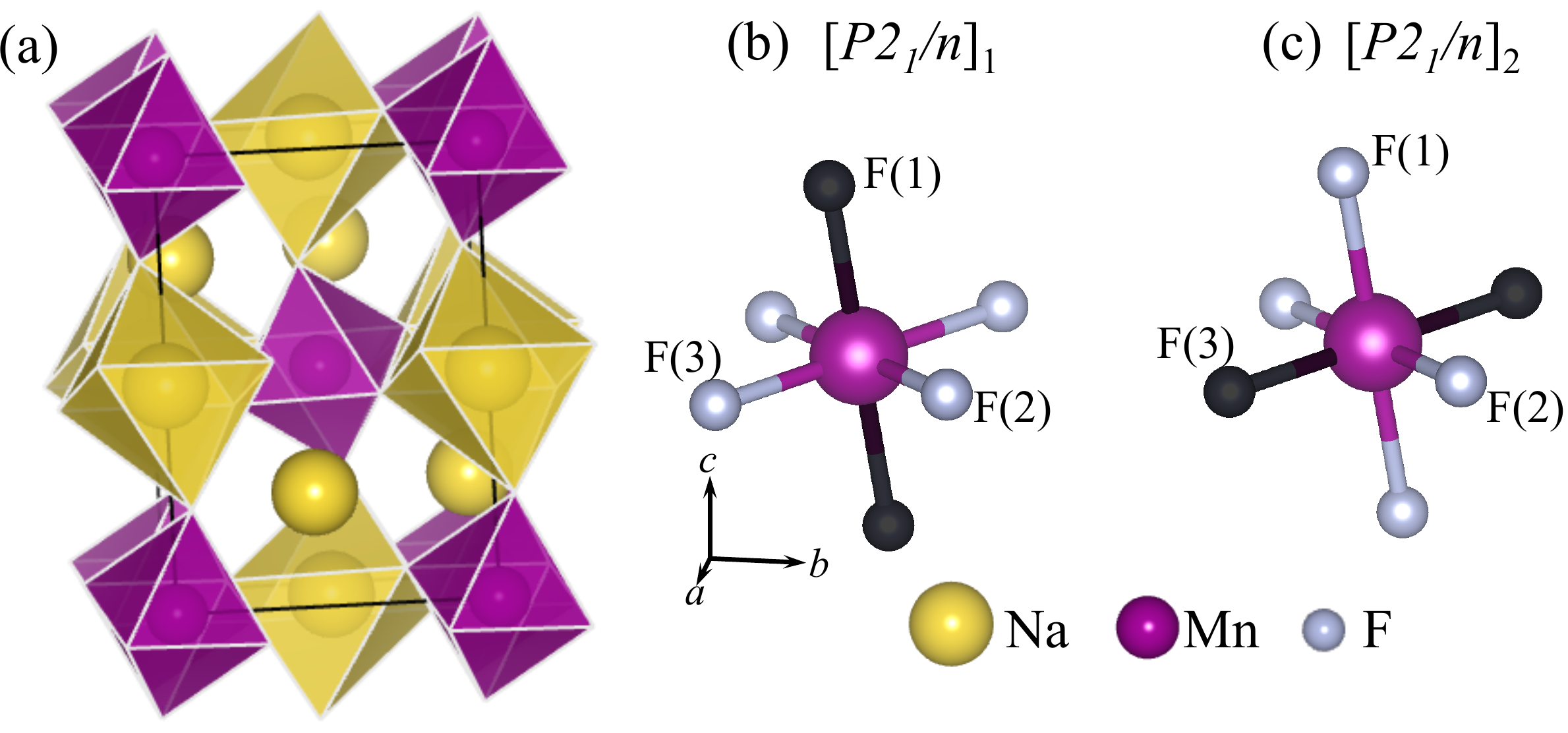}\vspace{-7pt}
\caption{%
Equilibrium zero kelvin
structure of Na$_3$MnF$_6$, and the 
MnF$_6$ octahedra of the (b) low pressure  \plow and (c) high pressure \phigh, 
phases.
The long Mn--F bond axis, referred in the text as the Jahn-Teller (JT) bond 
axis, is highlighted in black and the ferrodistortive packing of these 
octahedra along the $c$-axis is the distinguishing structural feature across the 
monoclinic-to-monoclinic transition.
} 
\label{fig:na3mnf6_p21n}
\end{figure}

Here we focus on the cryolite \nmf ($t = 0.89$), which 
possesses a strong first-order active Jahn-Teller $d^4$ Mn$^{3+}$ cation
and exhibits a distorted monoclinic $P2_1/n$ (space group no.\ 14) structure 
with both in-phase and out-of-phase MnF$_6$ octahedral rotations 
in its ground state (\autoref{fig:na3mnf6_p21n}).\cite{Englich:al0476} 
Interestingly, \nmf is reported to experimentally undergo an  
isostructural first-order monoclinc-to-monoclinic phase transition with hydrostatic pressure\cite{Carlson/Norrestam:1998}.
Such iso-structural/iso-symmetric phase transitions (IPT) 
transitions without symmetry reductions\cite{scott2010iso}  
are scientifically interesting, because they infrequently occur in 
inorganic condensed matter phases that are known to exhibit displacive 
transitions \cite{Rondinelli/Coh:2011}. 
Microscopic origins for such transitions 
include spin-state transitions,\cite{Christy:1995} 
electronic or magnetic polarization rotation,\cite{gibbs2011high} 
or from bond reconfigurations due to mechanical epitaxial strain constraints.\cite{chen2011low, christen2011stress, gou2013predicted, hatt2010strain} 
The IPT in \nmf is experimentally found to be reversible while 
showing characteristic first order hysteric behavior. 
Above and below the critical pressure value for the phase transition, 
the authors of Ref.~\onlinecite{Carlson/Norrestam:1998} were able to identify both 
changes in the amplitude of the short and long Mn--F bond lengths and the relative 
orientation of these bond pairs with respect to the monoclinic axes.
The atomic and electronic features responsible for the phase transition, however, remain 
unknown: is it due to spontaneous strain, the Mn electronic degrees of freedom, or 
strong electron--lattice interactions?

Herein, we use density functional theory calculations 
to identify the microscopic origin of the pressure-induced 
IPT in the fluoromanganate Na$_3$MnF$_6$. 
We find that the ground state monoclinic structure is stabilized through a combination 
of energy lowering MnF$_6$ octahedral rotations and Jahn-Teller (JT) bond 
elongations; the latter arise to remove the orbital degeneracy owing to 
electron occupation of the majority spin $e_g^\uparrow$-manifold presented by a 
cubic crystal field. 
These combined effects produce a crystal structure at ambient 
pressure that exhibits elongated Mn--F bonds along the long, 
crystallographic $c$ axis depicted in \autoref{fig:na3mnf6_p21n}(b), and is 
common for manganates with axial ratios $c/a>1$. \cite{kugel1982jahn, goodenough1961relationship}
Upon application of hydrostatic pressure in \nmf, we find 
a reorientation of the long JT Mn--F bond occurs 
for unit cell volumes of $240\pm0.1$\,~\AA${^3}$, 
corresponding to an experimental pressure between 
2.10~GPa and 2.15~GPa in agreement with Ref.~\onlinecite{Carlson/Norrestam:1998}.
At the electronic structure level, we find that the low-pressure fluoromanganate 
phases have fully occupied $d_{z^2-r^2}$ orbitals, whereas 
in the high-pressure phase, the occupancy reverses and is 
predominately of $d_{x^2-y^2}$ orbital character. 
Furthermore, we find that the pressure renormalizes the 
mode stiffness of the Jahn-Teller distortion, which due to the electron-lattice coupling facilitates the electronic transition. 
Uncovering the microscopic origins of the iso-structural first order phase transition sheds light on the effect of stress on the magnetic, electronic and structural degrees of freedom in \nmf. Thus, improving our understanding of electronic and structural transitions in bulk $d^4$ fluoromanganites, 
which may facilitate the design of IPTs in thin film geometries for future device applications.

\section{Computational \& Simulation Details}
We performed density functional theory calculations with the 
general gradient approximation (GGA) of Perdew-Burke-Ernzerhof 
revised for solids\cite{PBEsol:2008} (PBEsol) as implemented in the Vienna
{\it Ab initio} Simulation Package ({\sc vasp}) 
\cite{Kresse/Furthmuller:1996a,Kresse/Joubert:1999} 
with the projector augmented wave (PAW) method \cite{Blochl:1994} 
to treat the core and valence electrons using 
the following valence configurations:  
$2p^63s^1$ for Na,
$3p^64s^13d^6$ for Mn,
and $2s^22p^5$ for F.
We used a $7\times7\times7$
Monkhorst-Pack $k$-point mesh \cite{Monkhorst/Pack:1976} with Gaussian 
smearing (0.1~eV width) for the Brillouin zone (BZ) integrations and a 
600~eV plane wave cutoff. 
Spin-polarized structural relaxations 
were performed until the Hellmann-Feynman forces were less than 1
meV~\AA$^{-1}$ for each atom position.

\textit{Magnetic Order.$\quad$}%
The long-range spin order for \nmf has not been previously reported. 
Therefore, prior to carrying out the pressure study, we explored 
the energetics for A-type antiferromagnetic order 
(in-plane ferromagnetic coupling between Mn sites, with out-of-plane 
antiferromagnetic coupling) and ferromagnetic order on 
the monoclinic $P2_1/n$ (ambient) structure with the JT Mn--F 
bond directed along the $c$ axis.
These two magnetic configurations are compatible with the 20 atom unit cell 
used in all calculations.
From our total energy calculations of the fully relaxed phases, 
we find that the ferromagnetic configuration 
is essentially degenerate with the antiferromagnetic configuration: 
There is less than a  0.1\% difference in cell volume and an energy difference of 
0.4\,meV per formula unit (f.u.). 
These small differences between the magnetic variants also persist 
for calculations performed at elevated pressures.
Owing to the strong ionic character of the fluoride and the fact that there 
appears to be very weak spin--lattice coupling, we use structures with 
ferromagnetic (FM) order throughout to perform the hydrostatic pressure simulations.
As described below, the FM order yields excellent agreement with 
experimental structural data.
\textit{Application of Hydrostatic Pressure.$\quad$} %
We computationally mimic the experimental hydrostatic pressure  study 
by imposing the lattice constants and 
monoclinic angles reported in Ref.~\onlinecite{Carlson/Norrestam:1998} while 
allowing the internal coordinates to fully relax to obtain 
the total energies for both the low-pressure ([$P2_1/n$]$_1$)
and high-pressure ([$P2_1/n$]$_2$) phases.
Throughout, we distinguish between the two phases by using the 
space group label with an additional index, 1 or 2, appended to the end to indicate 
if it is the low-pressure (phase 1) or high-pressure (phase 2) \nmf structure.
The starting atomic configuration for the 
\plow phases employ the positions obtained from our 
fully relaxed zero pressure DFT-PBEsol simulations, whereas for the 
\phigh structural relaxations, we initialize the atomic positions to 
those reported experimentally.\cite{Carlson/Norrestam:1998} 
\section{Results}
\begingroup
\begin{table}
\begin{ruledtabular}
\centering
\caption{\label{tab:internalcomparison}
Comparison between the internal coordinates given in reduced 
units  and lattice parameters $a$, $b$, and $c$ (\AA) 
for monoclinic \nmf~($P2_1/n$) obtained from DFT-PBEsol calculations and  
x-ray synchrotron experiments.\cite{Englich:al0476}
The monoclinic angle, $\beta = 88.96^\circ$, is constrained to the experimental 
value.
The unit cell volumes obtained from DFT and experiment 
are 251.09 and 250.97\,\AA$^{3}$, respectively. 
The Mn and Na(1) cations occupy the Wyckoff positions (WP) $2a\, (0,0,0)$ and $2b\,  (0,0,1/2)$, respectively 
without free parameters, while all other atoms are free to 
displace.} 
\begin{tabular}{lcccc}
Atom &WP	& 	& PBEsol    &Experiment\cite{Englich:al0476} \\
\hline
Na(2) &$4e$	& $x$		& 0.510		& 0.509 \\
	&	& $y$		& 0.058		& 0.055 \\
	&	& $z$		& 0.750		& 0.750 \\
F(1) &$4e$	& $x$		& 0.122		& 0.115 \\
	&	& $y$		& -0.063	& -0.058 \\
	&	& $z$		& 0.761		& 0.767 \\
F(2) &$4e$	& $x$		& 0.722		& 0.719 \\
	&	& $y$		& 0.828		& 0.829 \\
	&	& $z$		& -0.058	& -0.053 \\
F(3) &$4e$	& $x$		& 0.162		& 0.163 \\
	&	& $y$		& 0.724		& 0.721 \\
	&	& $z$		& 0.068    	& 0.067 \\
\hline
$a$	&	&		& 5.472	   & 5.471 \\
$b$	&	&		& 5.684       & 5.683 \\
$c$	&	&		& 8.075       & 8.073 \\
\end{tabular}
\end{ruledtabular}
\end{table}
\endgroup

\subsection{Na$_3$MnF$_6$ Equilibrium Structure}
Before performing the hydrostatic pressure study on \nmf, 
we first determine the equilibrium structure at zero pressure 
with ferromagntic spin order. 
Consistent with experiments, we find a distorted monoclinic  
structure with in-phase MnF$_6$ octahedral rotations about the 
$c$ axis and out-of-phase tilts. The combination of rotations 
and tilts gives the $a^-a^-c^+$ tilt pattern as 
described within Glazer notation \cite{Glazer:1972}.
\autoref{tab:internalcomparison} shows that the PBEsol functional 
provides an accurate description of the structural parameters of 
\nmf when compared to experiment. The error in the cell volume is 
$\sim$0.05\%, under the constraint that the PBEsol structure 
has the same monoclinic angle as that reported in Ref.~\onlinecite{Englich:al0476}. 

\begin{figure}
\centering\includegraphics[width=0.88\columnwidth]{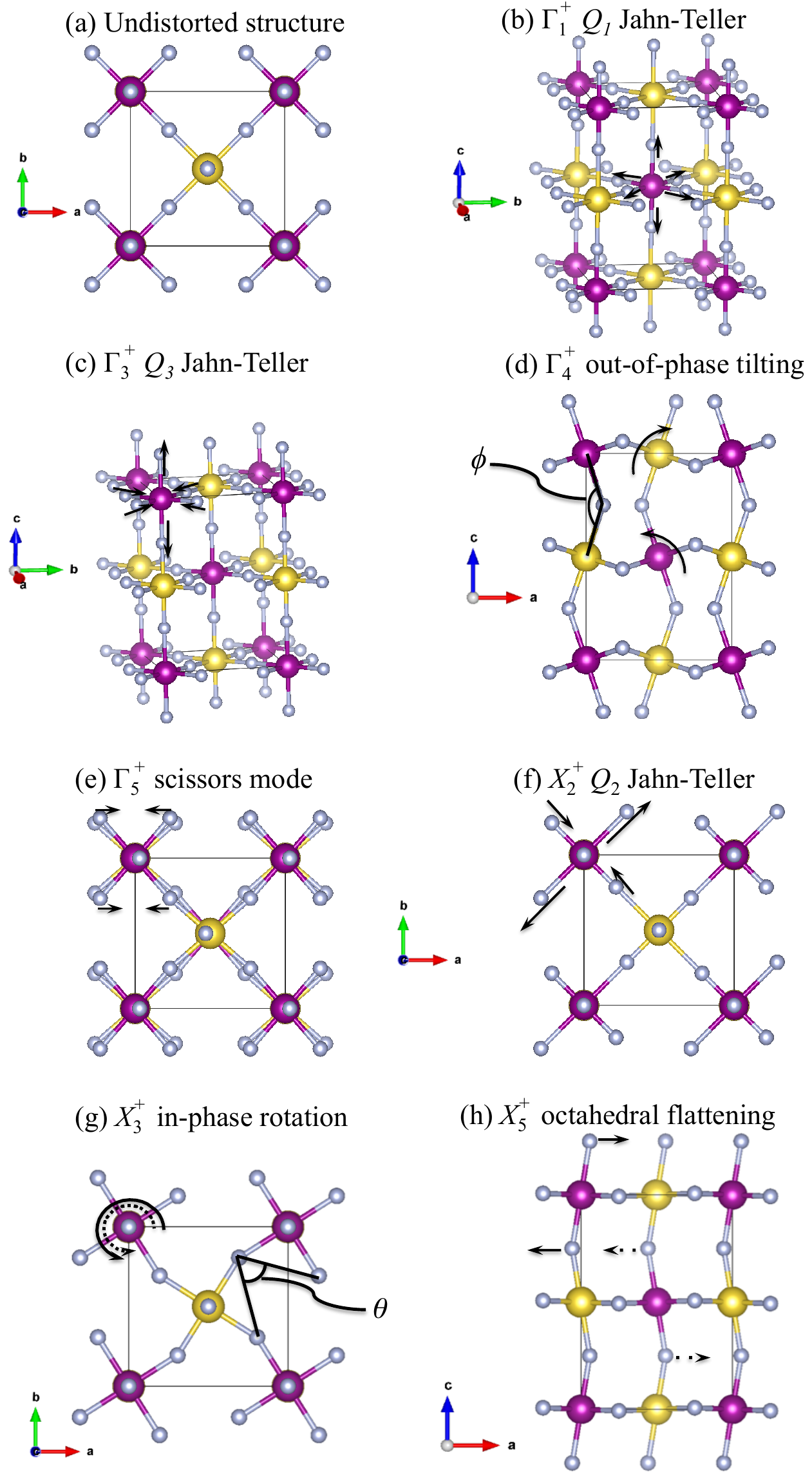}\vspace{-0.75\baselineskip}
\caption{%
Illustration of symmetry-adapted mode displacements which appear in 
monoclinic \nmf labeled according to irreps of the undistorted (a) $Fm\bar{3}m$ 
structure: 
(b) $\Gamma_1^+$: the $Q_1$ Jahn-Teller (JT) mode commonly 
referred to as the breathing distortion. 
(c) $\Gamma_3^+$:  the $Q_3$ (two out--four in) JT vibrational mode. 
(d) $\Gamma_4^+$: the out-of-phase tilting of octahedra about the $b$ axis. 
The tilt angle is measured as ($180^{\circ} - \phi$)/2.
(e) $\Gamma_5^+$: ``scissor'' mode which brings equatorial F anions in the 
MnF$_6$ octahedra closer together and reducing the intra-octahedral F--Mn--F 
bond angle from the ideal 90$^\circ$. 
(f) $X_2^+$: the $Q_2$ (two in--two out) bond stretching mode in the $ab$ plane. 
(g) $X_3^+$: in-phase rotations of the octahedra about the $c$ axis. The rotation angle is measured as ($90^{\circ} - \theta$)/2.
(h) $X_5^+$: anti-parallel apical F displacements which lead to a flattening 
of the MnF$_6$ octahedra.
Note that the Na(2) cations are excluded for clarity throughout, and broken 
lines indicate atomic displacements in the layer below.
} 
\label{fig:distortion_description}
\end{figure}

To understand the complex atomic distortions in the 
monoclinic phase, we first recognize that 
a group--subgroup relationship exists 
between the ideal high-symmetry cubic structure without 
octahedral distortions or rotations, $Fm\bar{3}m$, and the 
$P2_1/n$ structure.
\footnote{The occupied Wyckoff sites for \nmf 
in the $Fm\bar{3}m$ space group are 
$4a$ for Mn, 
$4b$ for Na(1), 
$8c$ for Na(2), and
$24e$ for  F.} 
Mode crystallographic analyses \cite{Isodisplace:2006,Perez-Mato:2010} then 
enable us to determine the full set of unique symmetry-adapted 
mode displacements active in the $Fm\bar{3}m \rightarrow P2_1/n$ transition 
(see \autoref{fig:distortion_description} for schematic illustrations), 
described as irreducible representation (irreps) of $Fm\bar{3}m$. 
We find that the cubic-to-monoclinic symmetry reduction requires a 
combination of at least two of these modes; based on the physical character (\autoref{fig:distortion_description}) and mode amplitudes (\autoref{tab:equi_vals}) of the distortions appearing in the ground state structure, we 
deduce that the symmetry reduction is 
driven by out-of-phase octahedral rotations about the $b$ axis  and 
in-phase rotations about c-axis, \emph{i.e.}, the combination of $\Gamma_4^+\!\oplus\!X_3^+$.

\begin{table}[t]
\begin{ruledtabular}
\centering
\caption{\label{tab:equi_vals}
Symmetry-adapted mode amplitudes (in \AA) 
present in the equilibrium \nmf \plow phase 
obtained from DFT-PBEsol calculations 
compared to experiment\cite{Englich:al0476}.
The mode amplitudes are 
normalized with respect to the parent $Fm\bar{3}m$ structure. 
} 
\begin{tabular}{llcc}
Mode						&Space group (number)		& DFT-PBEsol    & Experiment \\
\hline
$\Gamma_1^+$ 				&$Fm\bar{3}m$ (225)		& 0.154    	   & 0.175 \\
$\Gamma_3^+$ 			    	&$I4/mmm$	(139)	& 0.160	   & 0.083 \\
$\Gamma_4^+$ 			        & $C2/m$ (12)			& 1.421	   & 1.351 \\
$\Gamma_5^+$ 		              	& $C2/m$	(12)		& 0.087	   & 0.087 \\
$X_2^+$ 				& $P4_2/mnm$	 (136)	& 0.056	   & 0.038 \\
$X_3^+$ 				& $P4/mnc$	(128)	& 0.892	   & 0.916 \\
$X_5^+$ 				&  $Pnnm$ 		(58) 	& 0.696	   & 0.656 \\
\end{tabular}
\end{ruledtabular}
\end{table}

Given that the irreps form a complete basis for which to perform the 
structure decomposition of the monoclinic phase relative to the cubic 
structure, we can use the mode amplitude to make a quantitative comparison 
of the internal atomic positions obtained from our relaxed PBEsol simulation 
and the experimental structure.
\autoref{tab:equi_vals} reveals that there is only a modest difference  
($<\!10$\% error) between the PBEsol structure and experiment 
for all mode distortions that are not characterized by distortions of the  
Jahn-Teller type  ($\Gamma_3^+$, $X_2^+$). 
For example, our PBEsol calculation underestimate the 
experimental rotation angle ($12.5^{\circ}$) by 3.3\%, whereas the experimental
tilt angle ($19.4^{\circ}$) is overestimated by 5.7\%.
Both values are within the expected error of 20\% for 
perovskite systems with $a^-a^-c^+$ tilt patterns as predicted 
with semi-local exchange correlation functionals \cite{garciabenchmark2012}.

\begin{table}[b]
\begin{ruledtabular}
\centering
\caption{\label{tab:bond_len}Comparison of the 
Mn--F bond lengths for the ground state \nmf \plow structure 
computed with the PBEsol functional to experiment.\cite{Englich:al0476}
All units in~\AA.} 
\begin{tabular}{lcc}
Bond				& DFT-PBEsol &Experiment \\
\hline
Mn-F(1)			& 2.084		& 2.017 \\
Mn-F(2)			& 1.863		& 1.862 \\
Mn-F(3)			& 1.876		& 1.897 \\
\end{tabular}
\end{ruledtabular}
\end{table}

In the case of JT modes, we find that PBEsol is less accurate in predicting  
the distortion amplitudes. 
Specifically, we find a modest 12\% underestimation for the $Q_1$ mode 
[\autoref{fig:distortion_description}(b)], but large overestimations of 
47\% and 92\% for the $Q_2$ and $Q_3$ modes 
[\autoref{fig:distortion_description}(f) and (c)], respectively.
Closer inspection of the MnF$_6$ octahedra shows that these overestimations 
only manifest as small changes in Mn-F bond lengths of order 
10$^{-2}$~\AA. (\autoref{tab:bond_len}). 
Our calculations find that the long Mn--F(1) bond is enhanced, while 
the medium length Mn--F(3) bond is shortened relative to experiment.
There is a negligible difference in the short Mn--F(2) bond  in the 
MnF$_6$ octahedra.

The Jahn-Teller bond length discrepancies found when 
comparing the DFT predicted values to experiment are 
common for GGA-level exchange-correlation functionals, and are usually 
corrected by employing the plus Hubbard $U$ 
correction \cite{Sawada_LaMnO_JT_1997, Leonov_2010_atomic_dis, Rabe_JT_LMO_2013}. 
Indeed, for \nmf we obtain small improvements in the Jahn-Teller bond 
distortions with the Hubbard $+U$ correction\cite{Dudarev/Zuo_et_al:2506} (data not shown), however this 
is accompanied by a decrease in the accuracy of the unit cell volume, which should be 
well-reproduced in order to accurately evolve the system under pressure.
Despite the discrepancies, 
the trend in the Jahn-Teller bond distortions obtained at the PBEsol level are in 
good agreement with the experimental bond lengths, \emph{i.e.}, 
Mn--F(1){\textgreater}Mn--F(3){\textgreater}Mn--F(2). 
Thus, we find that the PBEsol functional without the Hubbard correction is a suitable 
exchange-correlation functional to use to explore the origin of the pressure-induced 
isosymmetric transition.

\begin{figure}[b]
\centering
\includegraphics[width=0.9\columnwidth]{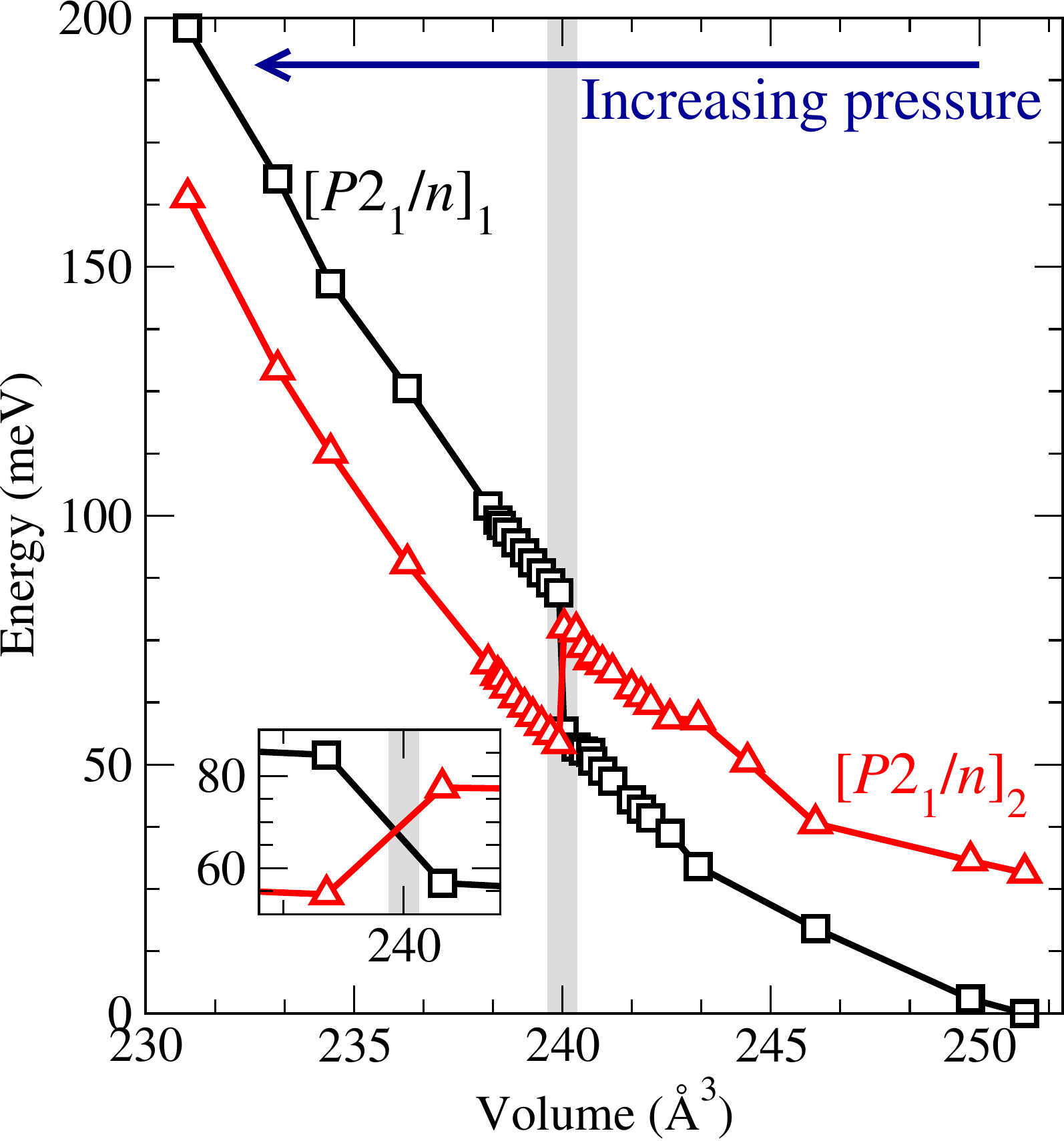}\vspace{-0.9\baselineskip}
\caption{%
Evolution of the total energy for the unique monoclinic 
\plow and \phigh structures with cell volume given relative to the 
relaxed zero kelvin \plow ground state structure. 
The shaded region highlights the pressure domain of phase coexistence and 
the transition point about which the stability of the monoclinic phases reverses. 
}
\label{fig:transtion_info}
\end{figure}

\subsection{Hydrostatic Pressure-induced Isosymmetric Transition}

We investigate phase stability of the monoclinic \nmf 
by computing the total energy at various volumes for both the \plow 
and \phigh structures.  
\footnote{
Fine sampling near the reported\cite{Carlson/Norrestam:1998} 
transition pressure is achieved by 
extrapolating best fit lines of the evolution of lattice parameters from 
the energetically more stable \phigh.} 
\autoref{fig:transtion_info} shows  that the energy of both monoclinic 
phases increase with increasing pressure (decreasing cell volume). 
At low-pressure \plow, exhibiting the Jahn-Teller bond axis oriented along the 
$c$ axis, is more stable than \phigh (JT bond axis in  the $ab$ plane) 
by $\sim$20\,meV.
This energetic stability persists until near a unit cell volume of 240\,\AA,  
where we find that both monoclinic structures are stable within our simulations. 
Upon further increase in pressure, \phigh is stabilized relative to \plow by nearly the 
same energetic difference.
We observe no change in space group symmetry 
or occupied Wyckoff positions for all volumes computed, 
consistent with available experimental results, 
which makes this transition isosymmetric. 

We now explore changes in the internal atomic positions with pressure, 
focusing on the cooperative behavior of the Mn--F octahedra. 
We first examine the changes in the Mn--F bond lengths.
Rather than using the symmetry-mode description described before, 
we parameterize the structural distortions to the MnF$_6$ octahedra in terms of 
first-order Jahn-Teller (JT) elongations $Q_2$ and $Q_3$ \cite{vanvleckJahnTeller}.
This allows us to describe the JT modes by the position of the surrounding ligands in an octahedral 
field whose normal coordinates are associated with the vibrational mode that leads to a crystalline 
field removing the orbital degeneracy. 
The $Q_2$ mode is a tetragonal ``two-in and two-out'' bond distortion in the 
$ab$ plane [$Q_2>0$, \autoref{fig:distortion_description}(c)], 
while $Q_3$ is a three-dimensional ``two-out and four-in'' bond distortion 
[$Q_3>0$, \autoref{fig:distortion_description}(f)]. 
In \nmf, the $Q_2$ and $Q_3$ vibrational modes are ferrodistortively aligned.
In the monoclinic systems studied here, the $Q_2>0$ mode elongates 
the Mn--F bonds that lie mainly along the $b$ axis and shortens those along $a$, 
while $Q_3>0$ shortens the Mn--F bonds in the $ab$ plane while elongating the bonds along $c$ 
axis.

\begin{figure}
\centering
\includegraphics[width=0.99\columnwidth]{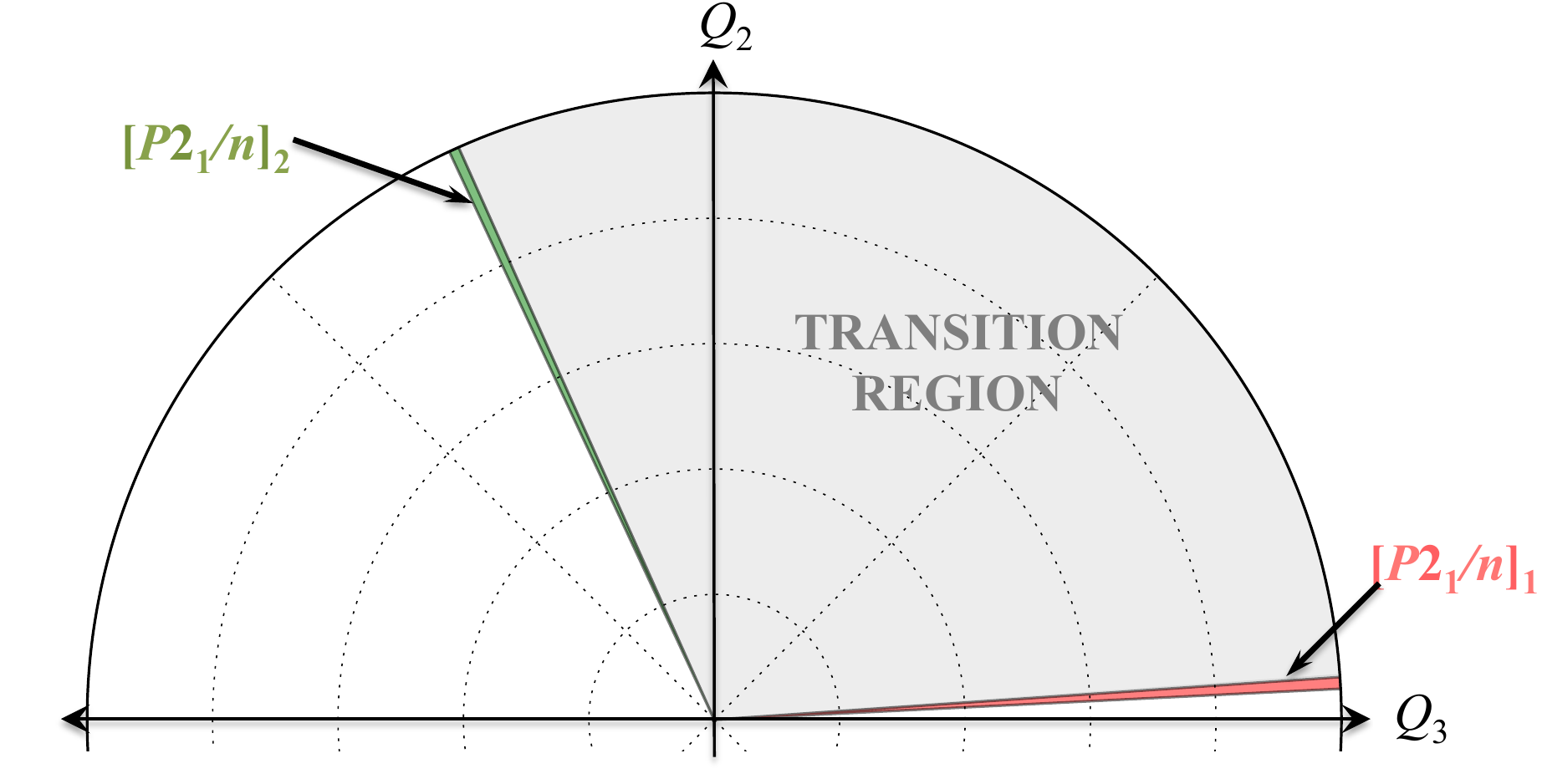}
\vspace{-1\baselineskip}
\caption{Phase stability of \plow and \phigh represented in the $Q_2-Q_3$ plane of 
the Jahn-Teller lattice distortions. 
The red, green, and gray wedges illustrate the stability regions 
for \plow, \phigh, and the transition region, revealing each structure is characterized 
by a dominant  $Q_2$ or $Q_3$  mode.
}
     \label{fig:Jahn_Teller_amo}
\end{figure}

We obtain the fraction of the $Q_2$ and $Q_3$ JT modes present 
in the stable monoclinic low- and high-pressure variants as a function of 
hydrostatic pressure by computing the magnitude of the Jahn-Teller normal modes as follows:
\begin{eqnarray*}
Q_2 &=& \frac{2}{\sqrt{2}} [\mathrm{F}(3)-\mathrm{F}(2)],\\\nonumber
Q_3 &=& \frac{2}{\sqrt{6}} [2\mathrm{F}(1)-F(2)-\mathrm{F}(3)]\,,\\\nonumber
\end{eqnarray*}
where, F(1), F(2), and F(3) are the Mn--F bond lengths directed 
along the $c$, $a$, and $b$ axes, respectively.
The JT phase is defined by $\Theta = \tan^{-1}Q_2/Q_3$. 
Within this formalism, $\Theta=0^{\circ}$ represents a JT distortion of
octahedra stretched along the $c$ axis.
Angles of $\Theta=\pm120^{\circ}$ correspond to 
JT distortions along the $b$ and $a$ axis, respectively \cite{goodenough1961relationship, kugel1982jahn, kanamori1960crystal}. 
\autoref{fig:Jahn_Teller_amo} shows the stability of the monoclinic \nmf 
structures in the $Q_2$-$Q_3$ plane with respect to  $\Theta$.
For \plow, $\Theta=2.99^{\circ}$ at ambient pressure and evolves to a maximum of 
value of $4.06^{\circ}$ at the critical pressure. 
Thus in the low pressure \nmf phase, the $Q_3$ JT mode dominates the 
$Q_2$ mode, leading to a tetragonal distortion of the MnF$_6$ octahedra that 
also elongates the $c$ axis and contracts the crystal in the $ab$ plane. 
Across the IPT there are no intermediate values of $\Theta\!$ (grayed region). 
In the high pressure phase \phigh, a drastic increase in the value 
of $\Theta$ occurs, \emph{i.e.}, $\Theta=114.1^{\circ}$ and reaches a maximum of $\Theta=115.0^{\circ}$, which 
indicates a first order transition to a $Q_2$-type JT distortion and is 
consistent with the long bond becoming the Mn--F(3) bond as observed in the \phigh phase.
\subsection{Jahn-Teller Electronic Effects}
We now investigate the electronic structure evolution across the IPT  
to understand the coupling between the electron and lattice degrees of 
freedom in \nmf.
Here, Mn(III) is found in a high-spin $d^4$ configuration with 
one electron in the $e_g$ manifold, which appears as a strong Jahn-Teller 
distortion in the octahedra. 
By definition the first-order Jahn-Teller (FOJT) effect of a transition metal cation in an 
octahedrally coordinated field acts to remove the electronic degeneracy of unevenly 
occupied $t_{2g}$ (weak JT) or $e_g$ (strong JT) orbitals by inducing a structural distortion \cite{carpenter2009symmetry,goodenough2004electronic,goodenough1998jahn}.
Thus, we anticipate that hydrostatic pressure directly affects the value of 
$\Theta$ in the $Q_2$-$Q_3$ subspace.
Importantly, the value of $\Theta$ also provides a measure of the average character 
of the occupied molecular orbital state $\Ket{\Theta}$ of the Jahn-Teller ion for a given 
static cooperative JT distortion as an arbitrary superposition of the $e_g$ orbitals wave 
functions.\cite{stroppa2013hybrid, kugel1982jahn} 
It can be approximated for a system at absolute zero as\cite{kanamori1960crystal}
\[
\Ket{\Theta} = \cos(\Theta/2)\Ket{d_{z^2-r^2}}+\sin(\Theta/2)\Ket{d_{x^2-y^2}}\,.
\]
We use Maximally Localized Wannier Functions (MLWF) \cite{MarzariVanderbilt1997, Mostofi_etal:2008} to transform our 
periodic wave functions forming the valence band (VB) and conduction band (CB) 
edges into a local real-space representation of the $e_g$ ($d_{x^2-y^2}$ and $d_{z^2-r^2}$) orbitals.
The subspace is  spanned by these atomic-like orbitals for the two Mn 
cations in the unit cell to evaluate the percent character of each orbital
contributing to $\Ket{\Theta}$. 
%

\begin{figure}
\centering
\includegraphics[width=0.98\columnwidth]{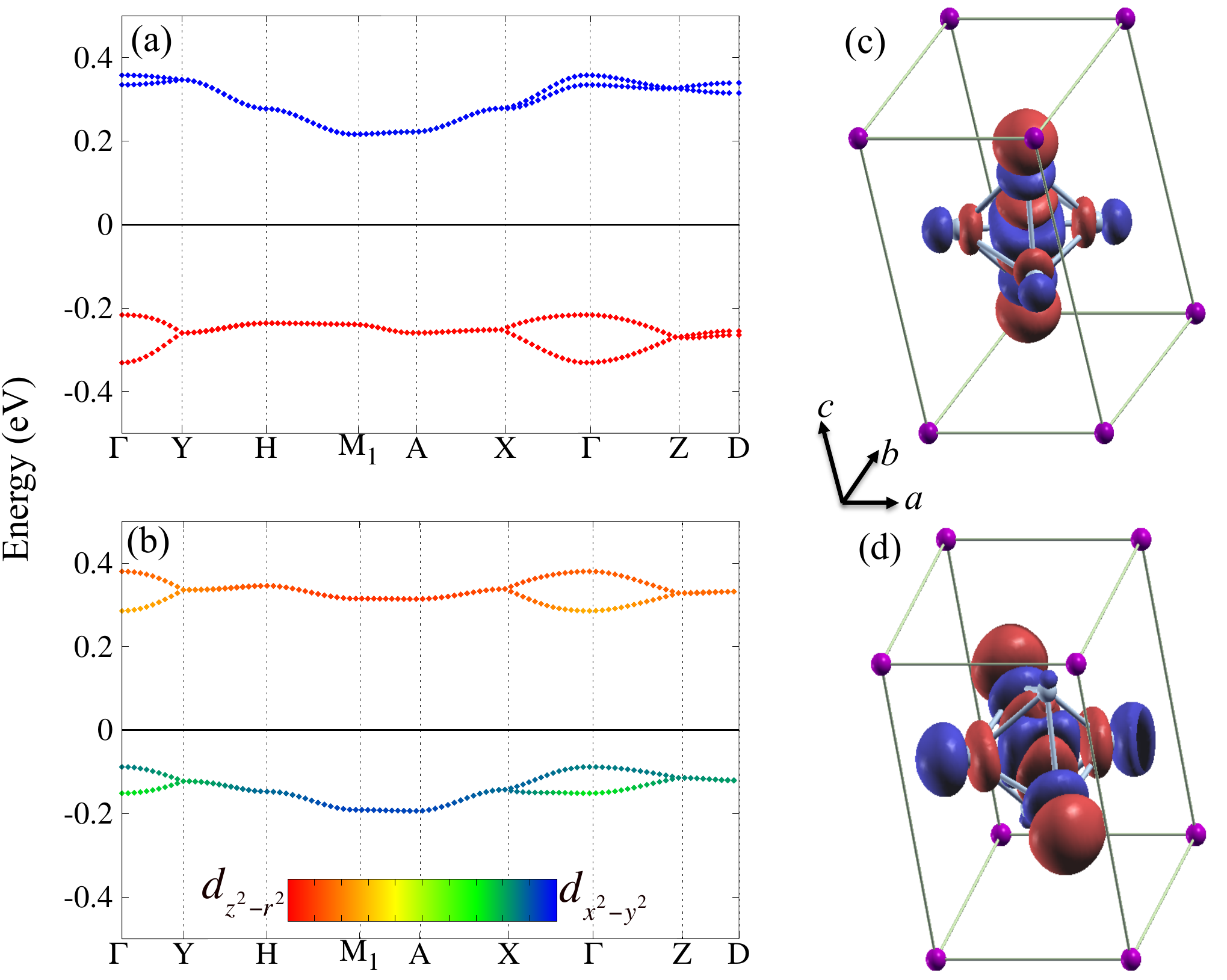}
\vspace{-0.7\baselineskip}
\caption{Low energy electronic band structure  about the Fermi level (horizontal line at 
0\,eV) with the projected $e_g$ orbital character for Na$_3$MnF$_6$ at 
(a) \plow (0.0\,GPA), and (b) \phigh (2.5\,GPa). 
The maximally-localized Wannier Functions (MLWF) for the occupied molecular 
orbital for the (c) \plow and (d) \phigh.
}
\label{fig:na3mnf6_DOS}
\end{figure}

\autoref{fig:na3mnf6_DOS}(a) shows the electronic band structure near 
the Fermi level ($E_F$) for \plow at ambient pressure. 
The occupied bands are largely 3$d_{z^2-r^2}$-like Mn states while 
the 3$d_{x^2-y^2}$ orbital forms the conduction band throughout the Brillouin zone (BZ). 
It is significant to note that $\Theta=2.99^{\circ}$ for \plow at ambient 
pressure and this corresponds to $\Ket{\Theta}$ being comprised of 99.9\% $d_{z^2-r^2}$ 
orbital character in agreement with the MLWF projection 
\autoref{fig:na3mnf6_DOS}(a).
This result is characteristic of the orbital splitting anticipated for $Q_3$\textgreater ~0, 
where the apical bonds in the MnF$_6$ octahedra are elongated and the equatorial bonds contract \cite{goodenough1961relationship}.
The MLWFs reveal strong localization of the atomic like 3$d_{z^2-r^2}$ orbital 
on the Mn cations below $E_F$ [\autoref{fig:na3mnf6_DOS}(c)];  the majority spin 
$e_g^\uparrow\!$ electron occupies the orbital directed along the long JT-bond to 
reduce the Coulombic repulsion induced by the F $2p$ electron cloud.

\autoref{fig:na3mnf6_DOS}(b) depicts the electronic structure for \phigh at 2.50~GPa.
From the projection onto the MLWFs, we observe 
in the high pressure phase that the VB and CB adopt mixed $e_g$-orbital 
character throughout the BZ.
The occupied orbital, although dominated by atomic $d_{x^2-y^2}$ character,  
has an admixture of $d_{z^2-r^2}$ [\autoref{fig:na3mnf6_DOS}(d)]. 
At 2.50\,GPa $\Theta=114.2^{\circ}$ in \phigh, and this angle predicts a wavefunction 
with 29.5\% ${d_{z^2-r^2}}$ orbital character. 
%
The mixed $e_g$ molecular orbital state is typical for manganates with a 
$Q_2$-like JT distortion, \emph{i.e.}, the long-JT bond axis is oriented in the 
$ab$ plane.\cite{goodenough1961relationship,medvedeva2002orbital,murakami1998resonant}

\begin{table}[b]
\begin{ruledtabular}\vspace*{-8pt}
\centering
\caption{\label{tab:high_equi_vals}
Symmetry-adapted mode amplitudes (in \AA) 
present in the  \nmf \phigh phase at 2.50~GPa
obtained from DFT-PBEsol calculations.
The mode amplitudes are 
normalized with respect to the parent $Fm\bar{3}m$ structure. 
} 
\begin{tabular}{llcc}
Mode						&Space group (number)		& DFT-PBEsol  \\
\hline
$\Gamma_1^+$ 				&$Fm\bar{3}m$ (225)		& 0.119    \\
$\Gamma_3^+$ 			    	&$I4/mmm$	(139)	& 0.103	   \\
$\Gamma_4^+$ 			        & $C2/m$ (12)			& 1.465	 \\
$\Gamma_5^+$ 		              	& $C2/m$	(12)		& 0.219	   \\
$X_2^+$ 				& $P4_2/mnm$	 (136)	& 0.083	   \\
$X_3^+$ 				& $P4/mnc$	(128)	& 0.892	   \\
$X_5^+$ 				&  $Pnnm$ 		(58) 	& 0.786	  \\
\end{tabular}
\end{ruledtabular}
\end{table}
\begin{figure}
\centering
\includegraphics[width=1\columnwidth]{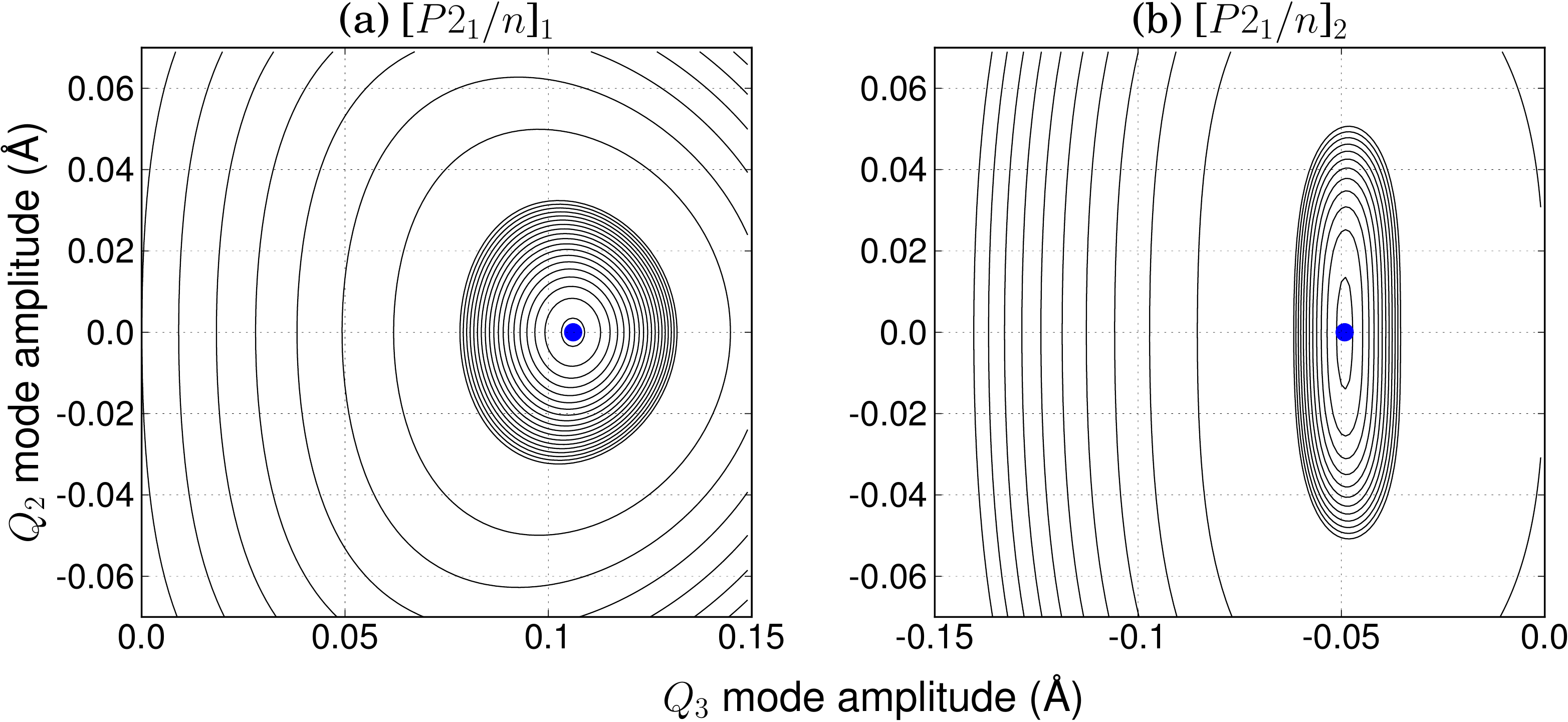}
\vspace{-1.6\baselineskip}
\caption{%
Calculated two-dimensional energy surface contours for 
(a) \plow at ambient pressure and (b) \phigh at 2.50 GPa with respect to the 
amplitude of the $Q_2$ and $Q_3$ JT modes derived 
from mode crystallography. 
The filled (blue) circles denote the position of minima in the potential energy surface.}
\label{fig:energy_surface}
\end{figure}

We now examine the energetics associated with the changes in $\Ket{\Theta}$ 
by computing the $Q_2-Q_3$ energy surface using the calculated total 
energies of  \plow and \phigh and fitting to a Taylor series expansion   
(\autoref{fig:energy_surface}). 
We achieve this by incrementally increasing the $Q_2$ ($X_2^+$) and $Q_3$ ($\Gamma_3^+$)  symmetry-adapted mode amplitudes in the \plow phase 
at ambient pressure (\autoref{tab:equi_vals}) and \phigh at 2.50~GPa (\autoref{tab:high_equi_vals}) with respect to the
undistorted reference structures.\footnote{The undistorted structures are constructed to have cell volumes and monoclinic angles that correspond to the pressure being studied 
with all mode amplitudes equal to zero relative to the DFT-PBEsol relaxed monoclinic structures. The energy maps are obtained from 225 self-consistent total energy calculations and then fitting these points to a polynomial expansion in orders of $Q_2$ and $Q_3$.}
In \autoref{fig:energy_surface}, mode amplitudes of $Q_2 > 0$ represent a JT distortion which elongates the Mn--F(3) bond and shortens the Mn--F(2) bonds in the $ab$ plane, while $Q_2 < 0$ shortens Mn--F(3) and elongates Mn--F(2). 
Similarly, $Q_3 > 0$ mode amplitudes correspond to a two-out--four-in JT distortion which stretches the apical Mn--F(1) while shortening the Mn--F(2) and Mn--F(3) equatorial bonds. Conversely, $Q_3 < 0$ gives the four-out--two-in JT bonding environment that elongates bonds in the $ab$ plane [Mn--F(2) and Mn--F(3)] and shortens the Mn--F(1) bond.

Considering only JT modes in the equilibrium volume of \plow at ambient pressure,  
we find a single energy minimum at $Q_3\sim0.11$ and $Q_2 = 0$ 
[\autoref{fig:energy_surface}(a)]. 
In this phase, a two-out--four-in $Q_3$ JT vibrational mode stabilizes \nmf at ambient pressure by approximately 38.6\,meV/f.u.\ over the undistorted phase, which is consistent with the 
structural and electronic investigations already described. 
Despite having a small finite amplitude of $Q_2$ ($X_2^+$) present in the equilibrium structure [\autoref{tab:equi_vals}], \autoref{fig:energy_surface}(a) indicates that in the absence of other distortions any nonzero amplitude of $Q_2$ leads to an 
energetic penalty in \plow.
Thus, an asymmetrical Mn--F bond stretching in the $ab$ plane is an 
unfavorable distortion at ambient conditions without the presence of another 
distortion. 
From \autoref{fig:energy_surface}(b), we observe that our two-dimensional energy surface predicts a single energy minimum for \phigh at 2.50~GPa. Here, the minimum is located at $Q_3\sim-0.048$ and $Q_2 = 0$. In the stable \phigh phase, it is noteworthy that the stability regime for the $Q_3$ vibrational mode is in a negative range. 
This indicates that unlike the \plow phase, at elevated pressures across the IPT a four-out--two-in $Q_3$  rather than two-out--four-in Mn--F distortion is required to stabilize the structure.
According to the JT theory, a $d^4$ cation with $Q_3 < 0$ distortion splits the $e_g$ degeneracy by occupying the $d_{x^2-y^2}$ orbital. Thus, the change in sign of the $Q_3$ mode is in accord with the $d_{x^2-y^2}$ dominate valence band edge observed in our electronic structure calculations (\autoref{fig:na3mnf6_DOS}). 
However, since $Q_2 = 0$ in the absence of any other distortion, at high 
pressure \nmf would exhibit four equivalent equatorial bonds, \emph{i.e.}, Mn--F(2) = Mn--F(3), in the reduced subspace. 
The latter is in contrast to the fully relaxed atomic and electronic structure of the 
\phigh phase, where the MnF$_6$ octahedra exhibit multiple distortions with an approximately 
$Q_2 > 0$-like JT bonding environment. 
Indeed, the symmetry-adapted mode amplitudes for \nmf at 2.50~GPa (\autoref{tab:high_equi_vals}) indicate that the $Q_2$ mode ($X_2^+$) is finite and 
larger in amplitude ($Q_2 = 0.083$) than at ambient pressure ($Q_2 = 0.056$). 

\begin{figure}
\centering
\includegraphics[width=0.99\columnwidth]{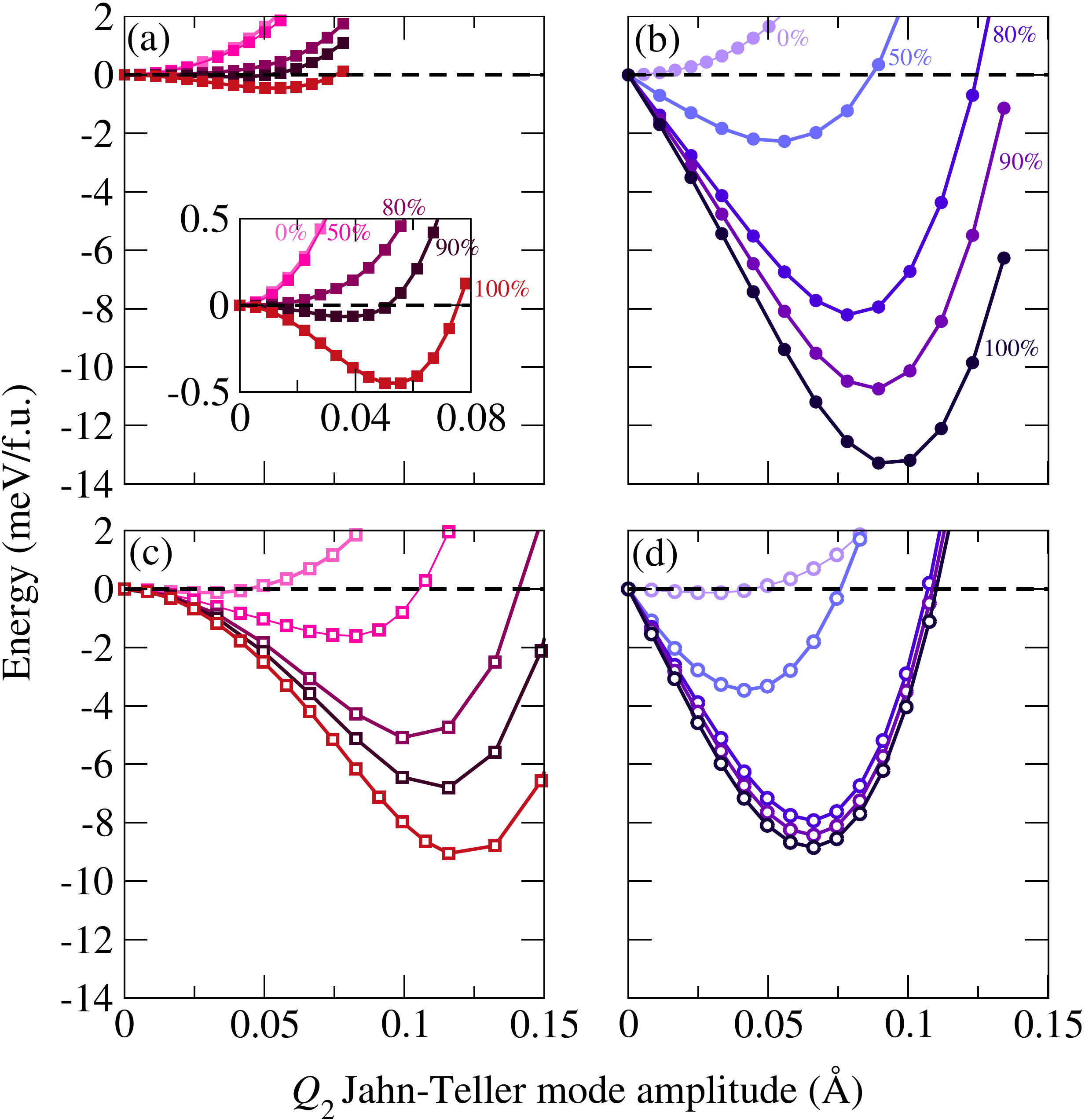}
\vspace{-1.7\baselineskip}
\caption{%
Energy evolution from the coupling of the $Q_2$ Jahn-Teller distortion to the tilt and rotation at (a) and (b) ambient pressure (filled symbols) and at (c) and (d) 2.50~GPa (unfilled symbols), respectively. Normalized energy gain obtained by increasing the amplitude of $Q_2$  at fixed percentages of the out-of-phase tilt found in the ground equilibrium structures. 
}
\label{fig:Landau_Q2_Q3}
\end{figure}

We now seek to identify which other structural distortion (non-Jahn-Teller mode) acts to stabilize nonzero $Q_2$ mode amplitudes in the equilibrium structures. 
From the mode decomposition analysis at ambient and elevated pressure, we observe that the largest contribution to the equilibrium structures arise from the tilt ($\Gamma_4^+$) and rotation ($X_3^+$) of the MnF$_6$ units. 
To understand the nature of the tilt--$Q_2$ and rotation--$Q_2$ coupling, we incrementally increase the mode amplitude of $Q_2$ at finite amplitudes of the tilt and rotation (0\%, 50\%, 80\%, 90\% and 100\% of their maximum) with respect to the undistorted reference structure.
\autoref{fig:Landau_Q2_Q3} shows that both the tilt and rotation modes 
couple to $Q_2$ and give a net energy gain in the free energy in both pressure regimes 
at finite amplitude.
From \autoref{fig:Landau_Q2_Q3}(a) we observe that  as the amplitude of tilt irrep increases in \plow, the curvature of the $Q_2$ free energy evolves from a positive parabola with a single minimum at the $Q_2 = 0$ to a double-well potential with  negative curvature about the origin and minima at finite $Q_2$ for tilt amplitudes $\geq\!$~80\%. This signifies a continuous softening of the $Q_2$ phonon mode induced by large amplitudes of the tilt mode ($\Gamma_4^+$ irrep), which effectively stabilizes the finite asymmetrical stretching observed in the ground state (\autoref{tab:equi_vals}) through a fourth-order biquadratic anharmonic interaction that renormalizes the mode stiffness of the  quadratic $Q_2^2$ mode. The behavior of the MnF$_6$ rotation--$Q_2$ coupling in  \autoref{fig:Landau_Q2_Q3}(b) shows that there is a large energetic gain associated with increasing amplitudes of the  rotation mode. The single minima of the parabolic energy curves indicate, however, that while this coupling contributes to the 
total energy it does not soften the $Q_2$ mode frequency, but rather shifts the mode amplitude of $Q_2$ to a non-zero value through a linear-quadratic interaction.

In the high pressure phase (2.50~GPa) [\autoref{fig:Landau_Q2_Q3}(c) and (d)], 
we observe that the $Q_2$ irrep is unstable with a small energy gain $<$ 1\,meV at  0\% tilt and rotation amplitudes. 
\autoref{fig:Landau_Q2_Q3}(c) shows that increasing the amplitude of the tilt distortion in 
the \phigh leads to an enhanced energy stabilization, which ultimately promotes 
a larger $Q_2$ distortion in the high pressure phase.  
In contrast, increasing the contribution of the rotational mode hardens the $Q_2$ mode [\autoref{fig:Landau_Q2_Q3}(d)], thereby reducing the energetic gain and leading to a smaller 
amplitude of $Q_2$ through the coupled $Q_2^2Q_{X_3^+}$ interaction.

\section{Discussion}

Our calculations and structural analysis indicate that the IPT 
is due to the spontaneous redistribution of the electronic charge density among 
the $e_g$ orbitals of the strong JT Mn cation. 
Specifically, we find that the low pressure system is stabilized with the long bond along the $c$ axis and  a filled $d_{z^2-r^2}$ orbital
which is characteristic of a $Q_3 > 0$ JT where $c/a > 1$ \cite{kugel1982jahn, goodenough1961relationship}. 
Mode crystallographic analyses coupled with phenomenological Landau investigations of the JT Mn--F bond distortions reveal that the lattice strain induced by hydrostatic pressure renormalizes the 
mode stiffness of the JT vibrational modes across the transition. 
The leading coefficient of the $Q_3^2$ term in the Hamiltonian changes sign 
under hydrostatic pressure.

Most significantly, we observe that across the transition the $Q_3$ JT irrep switches from a 
two-out-four-in ($Q_3 > 0$ ) in \plow to a 
four-out-two-in ($Q_3 < 0$) in the \phigh phase.
While the key structural signature defining \phigh is a $Q_2$-like Mn--F bonding arrangement, we contend that the primary factor in the stabilization of the high pressure phase in \nmf is the change in sign of the $Q_3$ distortion mediated by the 
$Q_2$--MnF$_6$ tilt interaction. \autoref{tab:equi_vals} and \autoref{tab:high_equi_vals} clearly show that the 
rotation amplitude  is unchanged under pressure. 
Our electronic and phenomenological  investigations indicate that at high pressure, it is more favorable for the charge to localize in the 
$d_{x^2-y^2}$-like orbital. 
Thus,  the characteristic $Q_2$-like Mn--F$_6$ bonding distortion that differentiates the \phigh and \plow phases is stabilized by second-order effects.
We show that a strong coupling exists between the tilt and $Q_2$ lattice degrees of freedom owing to an anharmonic interaction that stabilizes finite amplitudes of $Q_2$ in the equilibrium structures. The importance of this secondary effect is highlighted by comparing [\autoref{fig:Landau_Q2_Q3}(a) and (c)], as the energetic gain produced by the tilt--$Q_2$ coupling increases by approximately 20-fold in the \phigh phase at equilibrium over the ambient case. 

In addition, our electronic structure calculations show that hydrostatic pressure has an effect on the crystal field splitting in the stable phases of \nmf. We observe that energy gap between the filled $d_{z^2-r^2}$ and the unoccupied $d_{x^2-y^2}$ is reduced from 0.43~eV at ambient pressure to 0.34~eV at the transition. Across the critical pressure for the transition, the band gap is 0.37~eV at 2.20~GPa. The continued application of hydrostatic pressure in the high pressure phase further decreases the band gap at approximately 0.02\,eV/GPa which allows us to predict a insulator-to-metal transition around $\sim$20\,GPa.

\begin{figure}[t]
\centering
\includegraphics[width=0.91\columnwidth]{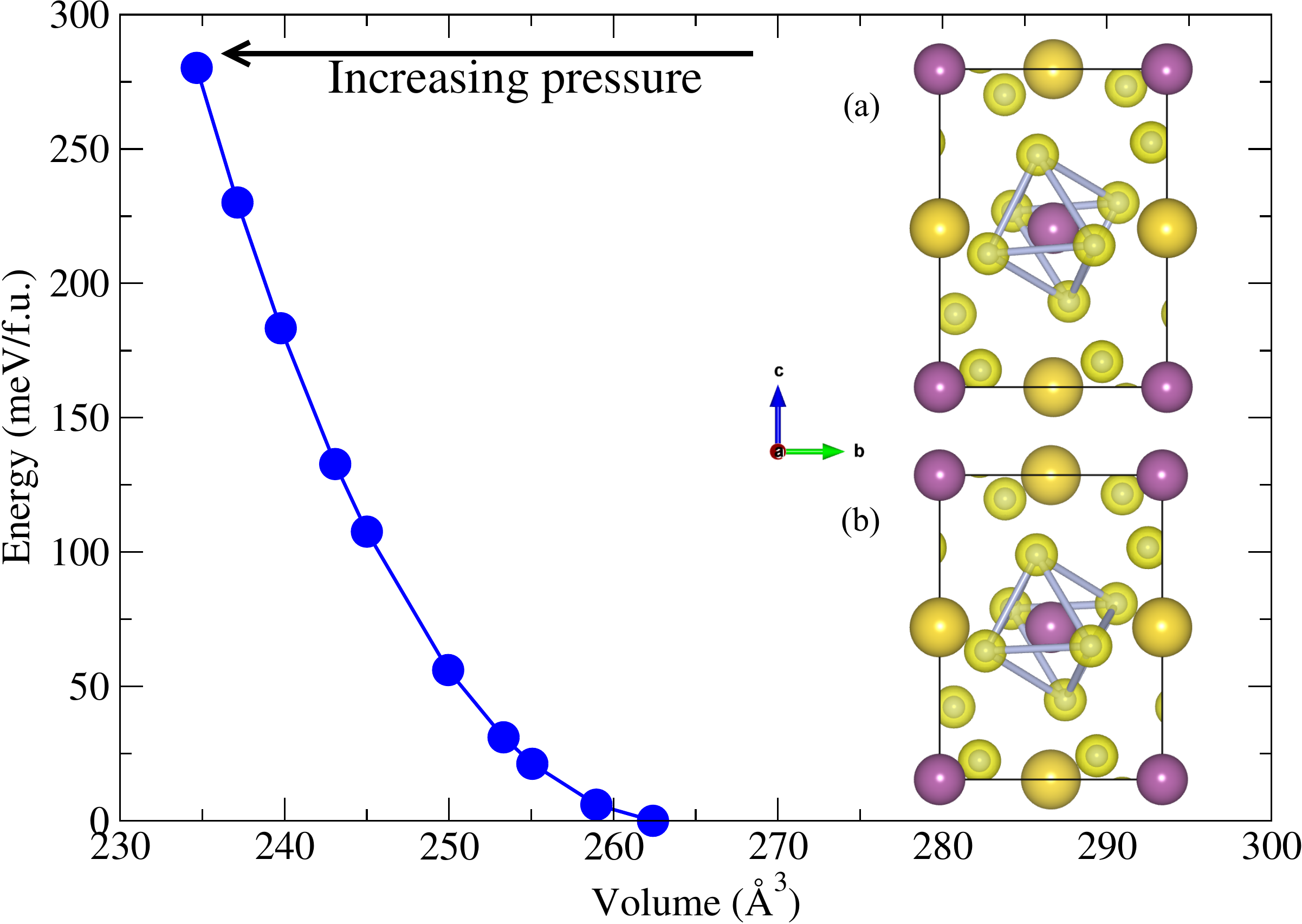}
\vspace{-0.9\baselineskip}
\caption{%
Evolution of the total energy of Na$_3$ScF$_6$ with cell volume relative to the DFT
relaxed ground state structure. 
The inset depicts the partial charge density obtained over the energy window ranging 
from -2 eV to $E_F$ at (a) ambient pressure (0.0~GPa) and (b) 6.82~GPa.
Unlike \nmf, neither a structural transition nor charge redistribution is observed.}
\label{fig:na3scf6_partial}
\end{figure}

Based on this understanding, we performed a similar set of calculations for the 
$d^0$ compound Na$_3$ScF$_6$, which also crystallizes with  $P2_1/n$ symmetry 
and for which experimental structural data under hydrostatic 
pressure exists.\cite{Carlson1998116}
We find there is no discontinuity in either the  total energies (\autoref{fig:na3scf6_partial}) 
or in the evolution of the Sc--F bond lengths up to a pressure of 6.82\,GPa, which 
indicates that no isosymmetric transitions occur up to a pressure that is 
approximately 4.7\,GPa higher than required for the 
fluoromanganate.
Indeed, \autoref{fig:na3scf6_partial} shows the partial charge density for Na$_3$ScF$_6$ 
obtained by integrating over a finite region from -2\,eV to $E_F$  is highly 
uniform at both ambient \autoref{fig:na3scf6_partial}(a) and high pressure \autoref{fig:na3scf6_partial}(b). 
The absence of both a JT instability and an anisotropic bonding environment is consistent  with the 
experimental findings and the proposed electronic origin for the IPT in \nmf, highlighting  the importance of the JT-active Mn(III).

\section{Conclusion}
We used first principles density functional calculations to study the 
electronic and atomic origins of the first-order pressure induced phase 
transition in \nmf.
We identified that the isosymmetric transition originates from 
two key features present in the fluoromangante: 
(i) a Jahn-Teller active Mn(III) ion with an $e_g$ orbital degeneracy, and 
(ii) the strong coupling of the cooperative Mn--F bond distortions to MnF$_6$ octahedral 
tilts with hydrostatic pressure.
We find that while simple structural arguments may identify the \plow phase of \nmf
as being dominated by a $Q_3$-type JT distortion and the \phigh phase as $Q_2$-type,  the combined effect of both distortions with collective MnF$_6$ tilts are essential to describing the stability of the structure at both ambient and under hydrostatic pressure.  
In the \plow structures the $Q_3 > 0$ JT distortion splits the $e_g$ manifold into atomic-like  occupied 
$d_{z^2-r^2}$ and an unoccupied $d_{x^2-y^2}$ orbitals, owing to 
an elongated JT Mn--F bond aligned along the crystallographic $c$-axis.
After the transition to the high-pressure phase \phigh, the JT-bond 
axis reorients into the $ab$ plane with the structure largely characterized by a $Q_2$-type bonding arrangement. 
The transition above 2.15~GPa is arguably driven by a spontaneous renormalization of the $Q_3$ vibrational mode, \emph{i.e.}, $Q_3 < 0$. 
The equilibrium structure of the \phigh phase is stabilized by secondary lattice effects which couple the tilt and $Q_2$ JT phonon modes to give the Mn--F(3) $>$ Mn--F(1) $>$ Mn--F(2), which is the signature of the high pressure phase of \nmf.

\begin{acknowledgments}
N.C.\ thanks Danilo Puggioni for useful discussions. 
N.C.\ and J.M.R.\ acknowledge the donors of The 
American Chemical Society Petroleum Research Fund for 
support (Grant No.\ 52138-DNI10).
DFT calculation were performed on the high-performance computing 
facilities available at the Center for Nanoscale Materials (CARBON Cluster) 
at Argonne National Laboratory, supported by the U.S.\ DOE, 
Office of Basic Energy Sciences (BES), DE-AC02-06CH11357.
\end{acknowledgments}


%
 
\end{document}